\documentclass[aps,prl,showpacs,twocolumn,superscriptaddress]{revtex4-1}
\usepackage{graphicx}
\usepackage[usenames]{color}
\usepackage{amssymb,amsmath}

\usepackage{array}

\newcommand{\eq}[1]{Eq.~(\ref{#1})}
\newcommand{\fig}[1]{Fig.~\ref{#1}}
\newcommand{\be}[1]{\begin{equation}\label{#1}}
\newcommand{\ee}{\end{equation}}
\usepackage[justification=centerlast,font=small]{caption}
\DeclareCaptionLabelSeparator{test}{ \textbf{$|$} }
\DeclareCaptionFont{small}{\small \bf}
\begin{document}

\title{Recollision as a probe of  magnetic field effects in non-sequential double ionization  }

\author{A. Emmanouilidou}
\address{Department of Physics and Astronomy, University College London, Gower Street, London WC1E 6BT, United Kingdom}
\author{T. Meltzer}
\address{Department of Physics and Astronomy, University College London, Gower Street, London WC1E 6BT, United Kingdom}
\begin{abstract}
Fully accounting for non-dipole effects in the electron dynamics, double ionization is studied for He driven by a near-infrared laser field and for Xe driven by a mid-infrared laser field. Using a three-dimensional semiclassical model, the   average  sum of the electron momenta along the propagation direction of the laser field is computed. This sum is found to be an order of magnitude larger than twice the average electron momentum along the propagation direction of the laser field in single ionization. 
Moreover, the average sum of the electron momenta in double ionization is found to be maximum at intensities smaller than the intensities  satisfying previously predicted criteria for the onset of magnetic field effects. It is shown that strong recollisions are the reason for this unexpectedly large value of the sum of the momenta along the direction of the magnetic component of the Lorentz force.

\end{abstract}
\pacs{}
\date{\today}

\maketitle
\section{Introduction}
Non-sequential  double ionization in two-electron atoms is a fundamental process that explores electron-electron correlation in strong fields.  As such, it has attracted a lot of interest in the field of light-matter interactions in recent years \cite{NSDI1,NSDI2}. The majority of theoretical studies on NSDI are delivered in the framework  of the dipole approximation, particularly 
the studies involving  the commonly used near-infrared laser fields and intensities \cite{Becker1}.  In the dipole approximation the vector potential $\mathrm{{\bf A}}$ of the laser field does not depend on space.  Therefore, magnetic field effects are neglected, since the magnetic field component of  the laser field  $\mathrm{{\bf B}=  {\bf \nabla}\times {\bf A}(t)}$ is zero. However, in the general case where $\mathrm{{\bf A}}$ depends both on space and time, an electron experiences  a Lorentz force   whose magnetic field component $\mathrm{{\bf F_{B}}}$ increases with increasing electron velocity, since $\mathrm{{\bf F_{B}}=q{\bf v}\times {\bf B}}$. It is important to account for magnetic field effects, since in strong field ionization high velocity electrons are often produced. Criteria for the onset of magnetic field effects both in the relativistic and  the non-relativistic limit  have already  been   formulated  \cite{Reiss1, Reiss2}. In the non-relativistic limit, where this work focuses,  magnetic field effects are expected to arise when the amplitude of the electron motion due to the magnetic field  component of the Lorentz force becomes 1 a.u., i.e.  $\mathrm{\beta_{0}=U_{p}/(2\omega c)\approx}$1 a.u.  \cite{Reiss1, Reiss2}, with U$\mathrm{_p}$ the ponderomotive energy.

Studies addressing magnetic field effects include using a 3D semiclassical rescattering model that accounts for $\mathrm{{\bf F_{B}}}$ to  successfully describe the observed ionization of Ne$\mathrm{^{n+}}$ ($\mathrm{n\leq8}$) in ultra-strong fields  \cite{Walker1}. Moreover, non-dipole effects  were addressed in theoretical studies of high-order harmonic generation, for instance,  by neglecting the Coulomb potential \cite{Joachain} or by using a first order expansion of the vector potential \cite{Brabec}. In recent studies of single ionization (SI), the electron momentum distribution along the propagation direction of the  laser field was computed using different quantum mechanical approaches   \cite{Corkum2,Corkum3,Drake, IvanovA}. For example, for H interacting with a 3400 nm laser field at intensities 0.5-1$\times$10$^{14}$ Wcm$^{-2}$ the average momentum along the propagation direction of the laser field was found to increase from 0.003 a.u. to 0.006 a.u. \cite{Corkum3}. Thus, for single ionization, the average of this momentum component    increases with increasing  $\mathrm{\beta_{0}}$ \cite{Corkum3, Corkum1}. If magnetic field effects are not accounted for,  then this momentum component averages to zero. The motivation for  these theoretical studies was a recent experimental observation of the average momentum in the propagation direction of the laser field  \cite{Corkum1}. 

\begin{figure} [ht]
\centering
 \includegraphics[clip,height=0.3\textwidth]{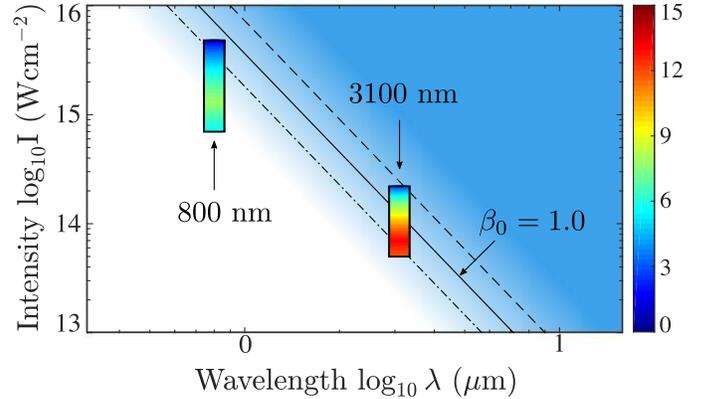}
 \caption{{\bf Range of validity of the dipole approximation and momentum in double ionization}. The white area indicates the range of intensities and wavelengths where the dipole approximation is valid. $\mathrm{\beta_{0}=0.5}$ a.u. (dot-dash line), $\mathrm{\beta_{0}=1}$ a.u. (solid line) and $\mathrm{\beta_{0}=2}$ a.u. (dash line). The arrows mark the 800 nm and  the 3100 nm wavelengths driving He and Xe, respectively. At these wavelengths, for a range of intensities, the color bars  indicate the ratio of the average sum of the electron momenta along the direction of $\mathrm{{\bf F_{B}}}$ for double ionization with twice the respective electron momentum for single ionization $\mathrm{\left <p_{y}^{1}+p_{y}^{2} \right >_{DI}/ (2\left <p_{y}\right>_{SI})}$. \hspace*{0.53cm}} 
\label{figure1}
\end{figure}
This work reveals another aspect of non-sequential double ionization (NSDI) which has not been previously addressed. The strong electron-electron correlation   in NSDI  is identified  as a probe of magnetic field effects both for near-infrared and mid-infrared intense laser fields. Specifically, the intensities considered are around 10$^{15}$ Wcm$^{-2}$ for He at 800 nm and around 10$^{14}$ Wcm$^{-2}$ for Xe at 3100 nm where the rescattering mechanism underlies double ionization \cite{Corkum}. 
For these intensities, it is found  that    the average sum of the two electron momenta along the propagation direction of the laser field is unexpectedly large.   It is roughly an order of magnitude larger than twice the  average of the respective electron momentum for single ionization. This average sum of the momenta for double ionization (DI) is shown to be  maximum at intensities smaller than the intensities satisfying  the criterion for the onset of magnetic field effects $\mathrm{\beta_{0}\approx}$1 a.u.  \cite{Reiss1, Reiss2}.   This is illustrated  in  \fig{figure1} for He driven by a near-infrared (800 nm) laser field and for Xe driven by a mid-infrared (3100 nm) laser field.  The motivation for choosing  near-infrared laser fields is that they are  very common in strong field studies. Mid-infrared laser-fields are chosen because magnetic field effects set in for small intensities, see \fig{figure1},   attracting a lot of interest in recent years  \cite{Keller, Biegert}.  

\section{Method}

 For the current  studies,  a 3D semiclassical model is employed that fully accounts for non-dipole effects during the time propagation.   For simplicity this model is referred to as 3D-SMND. It is an extension of a 3D semiclassical model that was previously formulated in the framework of the dipole approximation. This latter model is referred to as 3D-SMD. Thus, in the 3D-SMND model non-dipole effects are fully accounted for in the two-electron dynamics. Some of the successes of the 3D-SMD model are identifying  the mechanism responsible for the fingerlike structure \cite{Agapi1}, which  was predicted theoretically \cite{Taylor1} and was observed experimentally for He driven by 800 nm laser fields \cite{vshape1,vshape2}; investigating direct versus delayed pathways of NSDI for He driven by a 400 nm laser field while achieving excellent agreement with fully ab-initio quantum mechanical calculations \cite{Agapi2}; identifying the underlying mechanisms for the carrier-envelope phase effects observed experimentally in NSDI of Ar driven by an  800 nm   laser field  at a range of intensities \cite{Agapi3,Kling}.     The 3D-SMD model  is extended to the  3D-SMND  model employed in the current work to fully account for the magnetic field during time propagation. The Hamiltonian describing the interaction of the fixed nucleus two-electron atom with the laser field    is given by
\begin{equation}
\begin{aligned}
\mathrm{H=}&\mathrm{\frac{({\bf p}_{1}+{\bf A}(\mathrm{y_{1},t)})^2}{2}+\frac{({\bf p}_{2}+{\bf A}(\mathrm{y_{2},t}))^2}{2}-}\\
&\mathrm{-c_{1}\frac{Z}{|{\bf r}_{1}|}-c_{2}\frac{Z}{|{\bf r}_{2}|}+c_{3}\frac{1}{|{\bf r}_{1}-{\bf r}_{2}|}},
\end{aligned}
\label{eqn:1}
\end{equation}
where the vector potential $\mathrm{{\bf A}}$ is given by 
\begin{equation}
\mathrm{{\bf A}(y,t)=-\frac{E_{0}}{\omega}e^{-(\frac{ct-y}{c\tau})^2}\sin{(\omega t-ky)} \hat{x}},
\label{eqn:2}
\end{equation} 
 $\mathrm{\omega}$, k, E$_{0}$ are the frequency, wavenumber and strength of the electric component of the laser field,  respectively, and c is the velocity of light. $\mathrm{\tau=FWHM/\sqrt{ln4}}$ with FWHM the full width half maximum of the laser field. All Coulomb forces are accounted for by setting $\mathrm{c_{1}=c_{2}=c_{3}=1}$. In this work linearly polarized laser fields are considered. To switch-off a Coulomb interaction we set the appropriate constant equal to zero, for example, to switch-off the interaction of electron 1 with the nucleus we set $\mathrm{c_{1}=0}$.  For $\mathrm{\bf A}$ given by  \eq{eqn:2},  $\mathrm{\bf E}$ and $\mathrm{\bf B}$  are along the x- and z-axis, respectively, while  the propagation direction of the laser field and the direction of $\mathrm{\bf F_{B}}$ are along the y-axis.  Unless otherwise stated,  all Coulomb forces as well as the electric and the magnetic field are fully accounted for during time propagation. Moreover,  the Coulomb singularity  is addressed using  regularized coordinates \cite{KS}  which were also employed in the 3D-SMD  model \cite{Agapi1,Agapi2,Agapi3}. 

The initial state in the 3D-SMND model is taken to be the same as in the  3D-SMD  model  \cite{Agapi1,Agapi2,Agapi3}. It entails one electron tunneling through the field-lowered Coulomb potential with a non-relativistic quantum tunneling rate given by the Ammosov-Delone-Krainov (ADK) formula \cite{A1,A2}.  The momentum along the direction of the electric field is zero while the transverse one is given by a Gaussian distribution  \cite{A1,A2}. A non-relativistic ADK rate results in this Gaussian distribution  being centered around zero. In ref. \cite{Keitel}   non-dipole effects were accounted for in the ADK rate. It was shown  that the most probable transverse velocity ranges from 0.33 I$\mathrm{_{p}}$/c  to almost zero with increasing  $\mathrm{E_{0}/(2I_{p})^{3/2}}$, with I$\mathrm{_{p}}$ the ionization energy of the tunneling electron.  
In this work, the smallest intensities considered are 5$\times$10$^{13}$ Wcm$^{-2}$ for Xe and  7$\times$10$^{14}$ Wcm$^{-2}$ for He. At these intensities, if non-dipole effects are accounted for in the ADK rate,  the transverse velocity of the tunneling electron is centered around  0.17 I$\mathrm{_p^{Xe}}$/c for Xe which is 5.5$\times$10$^{-4}$ a.u.  (I$\mathrm{_{p}^{Xe}=}$0.446 a.u.) and 0.12 I$\mathrm{_{p}^{He}}$/c for He which is  7.9$\times$10$^{-4}$ a.u.  (I$\mathrm{_{p}^{He}=}$0.904 a.u.). These values are significantly smaller than the values of  the average momenta along the propagation direction of the laser field, which are presented in what follows. Thus, using the  non-relativistic ADK rate  is a good approximation for the quantities addressed in this work.
  The remaining electron is initially described by a microcanonical distribution \cite{Abrimes}. In what follows,  the tunneling and bound electron are denoted as electrons 1 and 2, respectively.  

\section{Results}

\subsection{$\mathrm{p_{y}}$ for single ionization of Xe and H}

The accuracy of the 3D-SMND model is established by computing the  momentum distribution along the propagation direction of the laser field, $\mathrm{p_{y}}$, for single ionization (SI)  and  by comparing it with available experimental and theoretical results. 
In ref. \cite{Keller}, the peak of the $\mathrm{p_{y}}$ distribution   was observed to shift in the direction opposite to the magnetic field component of the Lorentz force, $\mathrm{F_{\bf B}}$, for intensities  on the order of 10$^{13}$ Wcm$^{-2}$.  This shift was attributed to the combined effect  of the magnetic field and the Coulomb attraction of the nucleus \cite{Keller}. To compare with these experimental results, the shift of the peak of the $\mathrm{p_{y}}$ distribution is computed for Xe interacting with a 3400 nm and  a 44 fs FWHM laser field as the intensity increases from 3-6$\times$10$^{13}$ Wcm$^{-2}$. The shift of the peak of the $\mathrm{p_{y}}$ distribution is found to vary from  $\mathrm{-0.0055}$ a.u. to $\mathrm{-0.012}$ a.u.. These results are in agreement with the simulations and  experimental results presented in ref. \cite{Keller}. Moreover,  to compare with the results in ref. \cite{Corkum3}, the average of the  momentum  $\mathrm{\left <p_{y} \right >_{SI}}$ is computed for  H driven by a  3400 nm and a 16 fs FWHM laser field for intensities  0.5-1$\times$10$^{14}$ Wcm$^{-2}$. Using the 3D-SMND model, $\mathrm{\left <p_{y} \right >_{SI}}$   is found to vary from 0.0022 a.u. to 0.0046 a.u..  These values differ by 27\% from the results presented in  ref. \cite{Corkum3} and are thus in reasonable agreement. The difference may be due to non-dipole effects not accounted for in the ADK rate in the 3D-SMND model.   In addition, the quantum calculation used in ref. \cite{Corkum3} employs a 2D soft-core potential while a full 3D potential is employed by the 3D-SNMD model.
The single ionization results obtained in this work  were computed with at least 4$\times$10$^{5}$ events and therefore the statistical error introduced is very small.
\subsection{$\mathrm{\left<p_{y}\right>}$ for single ionization of He and Xe}

The 3D-SMND model is now employed to compute  $\mathrm{\left <p_{y} \right >_{SI}}$ for He driven by an 800 nm, 12 fs FWHM laser field and for  Xe driven by a 3100 nm, 44 fs FWHM laser field; the two laser fields  have roughly the same number of cycles.  First an analytic expression is obtained relating  $\mathrm{\left <p_{y} \right >_{SI}}$ with the average electron kinetic energy  $\mathrm{\left <E_{k} \right >_{SI}}$ \cite{Corkum1,Corkum2}. When an electron interacts with an electromagnetic field with all the Coulomb forces switched-off, i.e. $\mathrm{c_1=c_2=c_3=0}$ in \eq{eqn:1},     the equations of motion are $\mathrm{\dot{p}_y=-({\bf v}\times{\bf B})_y}$ and $\mathrm{\dot{p}_{x}=-({\bf v}\times{\bf B})_x-{E}}$. Keeping only first order terms in 1/c, $\mathrm{\dot{p}_{x}=- E}$ resulting in $\mathrm{p_y-p_{0,y}=p_{x}^2/(2c)-p_{0,x}^2/(2c)}$, with $\mathrm{p_{0,x/y}}$ the x/y components of the electron momentum at time $\mathrm{t_{0}}$.  The initial momentum of the tunneling electron along the electric field direction is set to zero, as in the 3D-SMND model, resulting in $\mathrm{p_y-p_{0,y}=p_{x}^2/(2c)=E_{k}/c}$. Thus, 
$\mathrm{\left <p_y \right >=\left<p_{0,y}\right>+\left<E_{k}\right>/c}$ is obtained. For this simple model $\mathrm{\left <E_{k} \right >}$ is the drift energy of the electron. As discussed in the Method section, if non-dipole effects are accounted for in the tunneling rate then $\mathrm{ \left<p_{0,y}\right>}$ varies from 0.33 I$\mathrm{_p}$/c to almost zero with increasing intensity.
In the 3D-SNMD model non-dipole effects are not included in the ADK rate and therefore  $\mathrm{ \left<p_{0,y}\right>}=0$. Indeed, using the 3D-SNMD model with $\mathrm{c_1=c_2=c_3=0}$, it is found that $\mathrm{\left <p_y \right >_{SI}=\left<E_k\right>_{SI}/c}$, see Table 1.

{\renewcommand{\arraystretch}{1.3}
\begin{table}[ht]
\centering
\begin{tabular}{rr c cc c cc c cc}
\toprule
\multicolumn{2}{c}{} & & \multicolumn{2}{c}{\textbf{SI} Z=2} & & \multicolumn{2}{c}{\textbf{SI}} & & \multicolumn{2}{c}{\textbf{SI} Z=2} \\
\multicolumn{2}{c}{} & & \multicolumn{2}{c}{$\mathrm{c_{1,2,3}}$=1} & & \multicolumn{2}{c}{$\mathrm{c_{1,2,3}}$=0} & & \multicolumn{2}{c}{$\mathrm{c_1}$=0,c$_\mathrm{2,3}$=1} \\ 
\cline{1-2} \cline{4-5} \cline{7-8} \cline{10-11} 
\multicolumn{2}{c}{ I $(\times10^{15}$Wcm$^{-2}$)} & & $\mathrm{\left\langle p_y \right\rangle}$ & $\mathrm{\left\langle E_k/c \right\rangle}$ & & $\mathrm{\left\langle p_y \right\rangle}$ & $\mathrm{\left\langle E_k/c \right\rangle}$ & & $\mathrm{\left\langle p_y \right\rangle}$ & $\mathrm{\left\langle E_k/c \right\rangle}$  $(\dagger)$ \\ 
\cline{1-2} \cline{4-5} \cline{7-8} \cline{10-11}
     & 0.7   \hspace{22pt} & & 3.6 & 2.3 & & 1.5 & 1.5 & & 2.0 & 2.2 \hspace{12pt} \\
He & 1.3   \hspace{22pt} & & 6.1 & 4.4 & & 3.4 & 3.4 & & 3.3 & 4.4 \hspace{12pt} \\
     & 4.8   \hspace{22pt} & & 31   & 31  & & 19 & 19 & & 19  & 21  \hspace{12pt} \\ \colrule
     & 0.05 \hspace{22pt} & & 3.2 & 1.6 & & 1.3 & 1.3 & & 1.8 & 1.6 \hspace{12pt} \\
Xe & 0.07 \hspace{22pt} & & 3.5 & 2.5 & & 2.1 & 2.1 & & 2.4 & 2.4 \hspace{12pt} \\
     & 0.22 \hspace{22pt} & & 11  & 12   & & 9.9 & 9.9 & & 9.1 & 11   \hspace{12pt} \\ 
\botrule
\multicolumn{11}{l}{\footnotesize{$(\dagger)$ Average momentum and kinetic energy given in $\times10^{-3}$ a.u.}}
\end{tabular}
\label{Table1}
\caption{{\bf Single ionization results for Xe and He.}\hspace*{0.5cm}}
\end{table}

  Next,  $\mathrm{\left <p_y \right >_{SI}}$ and $\mathrm{\left <E_k\right >_{SI}}$ are computed with the 3D-SMND model fully accounting for all Coulomb forces and the presence of the initially bound electron in driven He and Xe, i.e. $\mathrm{c_1=c_2=c_3=1}$ with $\mathrm{Z=2}$.  The tunneling electron  is the one that is mostly singly ionizing.   In \fig{figure2}, we show that, for He, $\mathrm{\left <p_y \right >_{SI}}$ varies from 0.0036 a.u. to 0.031 a.u. at intensities 0.7-4.8$\times$10$^{15}$ Wcm$^{-2}$. For Xe, $\mathrm{\left <p_y \right >_{SI}}$ varies from 0.0032 a.u. to 0.011 a.u. at intensities 0.5-2.2$\times$10$^{14}$ Wcm$^{-2}$.
  \begin{figure}
\centering
 \includegraphics[clip,height=0.33\textwidth]{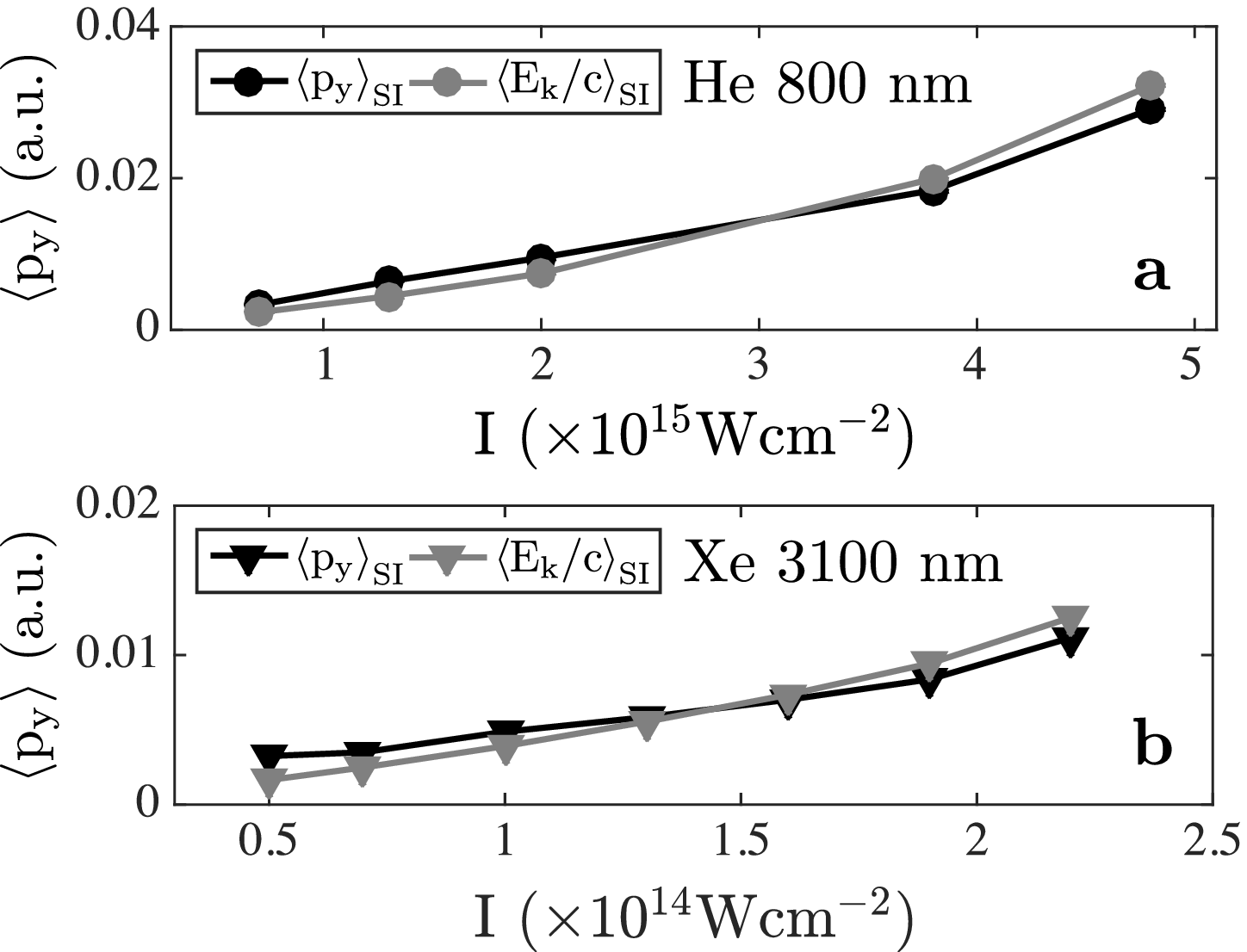}
 \caption{{\bf Single ionization of  He and Xe}.  $\mathrm{\left <p_y \right >_{SI}}$ and  $\mathrm{\left<E_k\right>_{SI}/c}$ are plotted as a function of intensity in (${\bf a}$) for He driven by an 800 nm laser field  and in (${\bf b}$)  for Xe driven by a 3100 nm laser field. \hspace*{5.7cm}}
\label{figure2}
\end{figure}
  In Table 1, it is shown  that $\mathrm{\left <p_y \right >_{SI}}$ and  $\mathrm{\left<E_k\right>_{SI}/c}$ when obtained with the full model do not differ by more than a factor of 3 from the values  obtained when  all Coulomb forces  are switched-off. Thus, the 
  simple model yields the correct order of magnitude for $\mathrm{\left <p_y \right >_{SI}}$. 
  It is also shown in Table 1, that with all Coulomb forces accounted for, $\mathrm{\left <p_y \right >_{SI}}$ is no longer equal to $\mathrm{\left<E_k\right>_{SI}/c}$ both for driven He and for driven Xe. For the full model, $\mathrm{\left<E_k\right>_{SI}}$ is no longer just the drift kinetic energy, mainly due to the interaction of the tunneling electron with the nucleus. Indeed, using the 3D-SMND model with this interaction switched off, i.e. $\mathrm{c_1=0}$ and  $\mathrm{c_2=c_3=1}$, $\mathrm{\left <p_y \right >_{SI}}$ is  roughly 
  equal to $\mathrm{\left<E_k\right>_{SI}/c}$, see Table 1.  $\mathrm{\left <p_y \right >_{SI}}$ is also shown in Table 1 to be  more sensitive than  $\mathrm{\left<E_k\right>_{SI}}$ to the  interaction of the tunneling electron with the nucleus. Summarizing the results for single ionization,  propagating classical trajectories with initial times determined by the ADK rate and all Coulomb forces switched-off yields the correct order of magnitude for  $\mathrm{\left <p_y \right >_{SI}}$.

\subsection{$\mathrm{\left <p_{y}^{1}+p_{y}^{2} \right >}$ for double ionization of He and Xe}

For double ionization, the average of the sum of the electron momenta along the propagation direction of the laser field  $\mathrm{\left <p_{y}^1+p_{y}^{2} \right >_{DI}}$, is computed for He driven by an 800 nm laser field and for Xe driven by a 3100 
nm laser field. The  parameters of the  laser fields are the same as the ones employed in the single ionization  section for He and Xe. The double ionization results obtained in this work  were computed with at least 2$\times$10$^{5}$ events and therefore the statistical error introduced is very small. The results are plotted in  \fig{figure3}{a}  for He at intensities 0.7-4.8$\times$10$^{15}$ Wcm$^{-2}$ and in  \fig{figure3}{b } for Xe at  intensities  0.5-2.2$\times$10$^{14}$ Wcm$^{-2}$. The values obtained for $
\mathrm{\left <p_{y}^{1}+p_{y}^{2} \right >_{DI}}$  are quite unexpected. Specifically,  $\mathrm{\left <p_{y}^{1}+p_{y}^{2} \right >_{DI}}$ is found to be roughly an order of magnitude larger than twice $\mathrm{\left <p_y \right >_{SI}}$, with $\mathrm{\left <p_y \right >_{SI}}$ computed in the previous section. For comparison, both $\mathrm{\left <p_{y}^{1}+p_{y}^{2} \right >_{DI}}$ and 2$\mathrm{\left <p_y \right >_{SI}}$ are 
displayed in \fig{figure3}. It is shown that $\mathrm{\left <p_{y}^{1}+p_{y}^2 \right >_{DI} \approx 8\times2\left <p_y \right >_{SI}}$ for He at 1.3$\times$10$^{15}$ Wcm$^{-2}$, while $\mathrm{\left <p_{y}^{1}+p_{y}^{2} \right >_{DI} \approx 13\times2\left <p_y 
\right >_{SI}}$ for Xe at 7$\times$10$^{13}$ Wcm$^{-2}$. For 1.3$\times$10$^{15}$ Wcm$^{-2}$ and 800 nm $\mathrm{\beta_{0}=0.18}$ a.u., while  for 7$\times$10$^{13}$ Wcm$^{-2}$  and 3100 nm $\mathrm{\beta_{0}=0.58}$ a.u.. Thus, $\mathrm{\left <p_{y}^{1}+p_{y}^{2} \right >_{DI}}/2\mathrm{\left <p_y \right >_{SI}}$ is found to be maximum at intensities    
 considerably smaller than the intensities  corresponding to $\mathrm{\beta_{0}\approx1}$ a.u., i.e.  the  criterion for the onset of  magnetic field effects \cite{Reiss1,Reiss2}. This is shown in  \fig{figure1}. 
 Moreover,  unlike $\mathrm{\left <p_y \right >_{SI}}$  which increases with increasing intensity as expected \cite{Corkum3},   $\mathrm{\left <p_{y}^{1}+p_{y}^{2} \right >_{DI}}$ after reaching a maximum decreases with increasing intensity for the range of intensities currently considered (\fig{figure3}a,b). What is the mechanism responsible for this pattern? We answer this question in what follows.

\begin{figure} [h]
\centering
 \includegraphics[clip,height=0.5\textwidth]{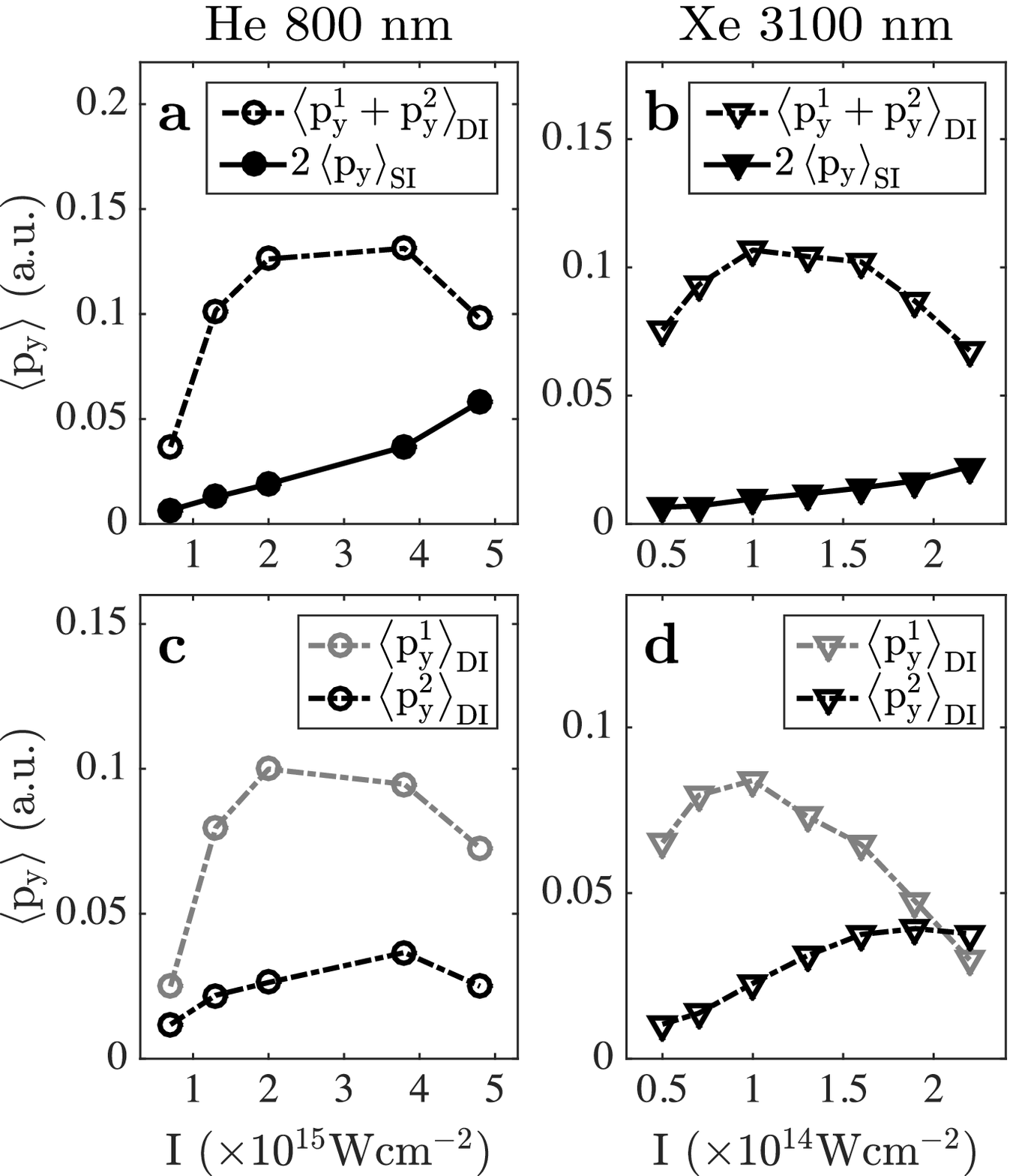}
 \caption{ {\bf Double ionization of He and Xe}. $\mathrm{\left <p_{y}^{1}+p_{y}^{2} \right >_{DI}}$ and 2$\mathrm{\left <p_y \right >_{SI}}$ are plotted as a function of intensity   in (${\bf a}$) for He  driven by an 800 nm laser field  and in   (${\bf b}$) for  Xe driven by a 3100 nm laser field.   $\mathrm{\left <p_{y}^{1}\right>_{DI}}
 $ and  $\mathrm{\left <p_{y}^{2}\right>_{DI}}$ are plotted as a function of intensity   in (${\bf c}$) for He  driven by an 800 nm laser field  and in   (${\bf d}$) for  Xe driven by a 3100 nm laser field. \hspace*{1cm} }
\label{figure3}
\end{figure}

  \subsection{Recollision probing magnetic field effects}

The average electron momentum along the propagation direction is non zero when the magnetic field component of the Lorentz force $\mathrm{{\bf F_{B}}}$ is accounted for. This force increases  with increasing intensity---increasing strength of the magnetic field---and with increasing velocity along the direction of the electric field. $\mathrm{\left <p_{y}^1+p_{y}^2 \right >_{DI}}$ is found to be  maximum at 1.3$\times$10$^{15}$ Wcm$^{-2}$ for  800 nm and at  7$\times$10$^{13}$ Wcm$^{-2}$  for 3100 nm, intensities  where the strength of the magnetic field is not large. It then follows that it must be the velocities of the two escaping electrons that are significantly larger at these intensities than at the higher intensities considered in this work. 
Large electron velocities at intermediate intensities are a result of  strong electron-electron correlation, i.e. of the rescattering mechanism \cite{Corkum}. In the rescattering scenario after electron 1 tunnels in the field-lowered Coulomb potential it accelerates in the strong laser field and can return to the core and undergo a collision with the remaining electron  \cite{Corkum}.  In what follows evidence is provided  that the large values of $\mathrm{\left <p_{y}^1+p_{y}^2 \right >_{DI}}$ are due to recollisions. Specifically,   
it is shown that recollisions are strong resulting in overall large electron velocities roughly at the intensities where  $\mathrm{\left <p_{y}^{1}+p_{y}^{2} \right >_{DI}}$ is maximum.  It is also shown that  recollisions are soft resulting in overall smaller velocities at higher intensities where    $\mathrm{\left <p_{y}^{1}+p_{y}^{2} \right >_{DI}}$ is found to be smaller.

This transition from strong to soft recollisions is demonstrated in the context  of He driven by an 800 nm laser field at intensities 0.7$\times$10$^{15}$ Wcm$^{-2}$, 2.0$\times$10$^{15}$ Wcm$^{-2}$ and 3.8$\times$10$^{15}$ Wcm$^{-2}$. To do so, an analysis of the doubly ionizing events is performed. In \fig{figure4}, the distribution of the tunneling and  recollision times is plotted. As expected, for the smaller intensities (\fig{figure4}a1,b1), electron 1 tunnel-ionizes at times around the extrema of the laser field.  For 3.8$\times$10$^{15}$ Wcm$^{-2}$ (\fig{figure4}c1) the electric field is sufficiently strong so that electron 1 can tunnel-ionize at times other than the extrema of the field. 
The distribution of the recollision times is also plotted. This time is identified for each double ionizing trajectory as the time  that the electron-electron potential energy $\mathrm{1/\left |{\bf r_{1}}-{\bf r_{2}}\right|}$ as a function of time is maximum.  
For the smaller intensities  the recollision times are   centered roughly around $\mathrm{\pm 2nT/3}$, with n an integer and T the period of the laser field, as expected from the rescattering model \cite{Corkum} (\fig{figure4}a2,b2). At  3.8$\times$10$^{15}$Wcm$^{-2}$  the recollision times shift and are centered around the extrema of the laser field (\fig{figure4}c2).  This shift of the recollision times  signals a transition from strong to soft recollisions \cite{Agapi4}. 
This transition is further corroborated by the average kinetic energy of each electron, $\mathrm{\left<E_{k}^{1,2}\right>}$,  
 plotted in \fig{figure4} as a function of time;  zero time is set equal to the recollision time of each double ionizing trajectory. For smaller intensities,  $\mathrm{\left<E_{k}^{1,2}\right>}$ changes sharply at the recollision time (\fig{figure4}a3,b3). 
 The change in $\mathrm{\left<E_{k}^{1,2}\right>}$ is much smaller at  3.8$\times$10$^{15}$ Wcm$^{-2}$ ( \fig{figure4}c3). The above results show that for the smaller intensities electron 1 tunnel-ionizes around the extrema of the field. It then returns to the core, roughly when the electric field is small, with large velocity and undergoes a recollision with electron 2 transferring a large amount of energy (strong recollision). The velocities of both electrons along the direction of the electric field are determined mainly by the vector potential at the recollision time. Thus,  both electrons escape mainly either parallel or antiparallel to the electric field. Indeed, this is the pattern seen in the plots of the   correlated momenta along the direction of the electric field  in \fig{figure4}a5,b5 where the highest  density is in the first and third quadrants. The correlated momenta are plotted in units of $\mathrm{\sqrt{2U_p}}$, with U$\mathrm{_p}$ the ponderomotive energy  equal to $\mathrm{E_{0}^2/(4\omega^2)}$. These patterns of the correlated momenta are consistent with direct  double ionization, that is, with both electrons ionizing shortly after  recollision takes place \cite{RESI2}. Indeed, analyzing the  double ionizing events it is found that for He  at 0.7$\times$10$^{15}$ Wcm$^{-2}$ direct  double ionization contributes 80\%.  Delayed double ionization events contribute 20\%. In delayed double ionization, also known as RESI  \cite{RESI1, RESI2},  one electron ionizes soon after recollision takes place, while the other electron ionizes with a delay \cite{Agapi2}.
 In contrast, at the higher intensity of   3.8$\times$10$^{15}$ Wcm$^{-2}$, electron 1 tunnel-ionizes after the extrema of the laser field. It then follows a short trajectory and returns to the core when the electric field is maximum with small velocity. Electron 1  transfers a small amount of energy to electron 2 (soft recollision). The velocities of electron 1 and 2 along the direction of the electric  field are determined mostly by the values of the vector potential at the tunneling and recollision  times, respectively. As a result, the two electrons can escape opposite to each other along the direction of the electric field.  This  pattern is indeed seen  in the plots of the correlated momenta  in \fig{figure4}c5 with high density in the second and fourth quadrants. This antiparallel pattern was predicted in the  context of strongly-driven N$_{2}$ with fixed nuclei \cite{Agapi4}.  It was also seen in the case of Ar driven by intense ultra-short laser fields \cite{Agapi3} in agreement with experiment \cite{Kling}.
  A similar analysis for Xe driven at 3100 nm is performed (not shown). Similar results are obtained. It is found that for driven Xe  strong recollisions transition to soft ones  roughly around 22$\times$10$^{13}$ Wcm$^{-2}$.

 \onecolumngrid
\begin{center}
\begin{table}[h]
\centering
\begin{tabular}{rrcccccccccc}
\toprule
\multicolumn{2}{c}{} & $\quad$ & \multicolumn{4}{c}{\textbf{NSDI} $\mathrm{c}_{\mathrm{1,2,3}}$=1, Z=2} & $\quad$ & \multicolumn{4}{c}{ \textbf{NSDI} $\mathrm{c}_{\mathrm{1,2,3}}$=0 } \\
\multicolumn{2}{c}{} & & \multicolumn{4}{c}{} & & \multicolumn{4}{c}{$\mathrm{t}_\mathrm{0}=\mathrm{t}_\mathrm{rec}$, $\mathrm{p}^\mathrm{1}_\mathrm{0}=\mathrm{p}^\mathrm{1}_\mathrm{rec}$, $\mathrm{p}^\mathrm{2}_\mathrm{0}=\mathrm{p}^\mathrm{2}_\mathrm{rec}$} \\ 
\cline{1-2} \cline{4-7} \cline{9-12}
\multicolumn{2}{r}{I $(\times10^{15}$Wcm$^{-2}$)} & & $\mathrm{\left\langle p_y^2 \right\rangle}$ & $\mathrm{\left\langle E^2_k/c \right\rangle}$  & $\mathrm{\left\langle p_y^1 \right\rangle}$ & $\mathrm{\left\langle E^1_k/c \right\rangle}$ & & $\mathrm{\left\langle p_y^2 \right\rangle}$ & $\mathrm{\left\langle E^2_k/c \right\rangle}$  & $\mathrm{\left\langle p_y^1 \right\rangle}$ & $\mathrm{\left\langle E^1_k/c \right\rangle}$   $(\dagger)$ \\
\cline{1-2} \cline{4-7} \cline{9-12}
 & 0.7  \hspace{20pt} & & 12 & 14 & 25 & 16 & & 13 & 15 & 25 & 16 \hspace{11pt} \\
He & 1.3 \hspace{20pt} & & 22 & 22 & 80 & 21 & & 24 & 24 & 82 & 20 \hspace{11pt} \\
 & 4.8 \hspace{20pt} & & 25 & 32 & 73 & 47 & & 33 & 37 & 74 & 43 \hspace{11pt} \\
\colrule
 & 0.05 \hspace{20pt} & & 10 & 11 & 65 & 9 & & 11 & 11 & 67 & 8 \hspace{11pt} \\
Xe & 0.07 \hspace{20pt} & & 14 & 13 & 80 & 10 & & 15 & 14 & 81 & 9 \hspace{11pt} \\
 & 0.22 \hspace{20pt} & & 38 & 27 & 30 & 17 & & 45 & 34 & 30 & 16 \hspace{11pt} \\ 
 \botrule
 \multicolumn{10}{l}{\footnotesize{$(\dagger)$ Average momentum and kinetic energy given in $\times10^{-3}$ a.u.}}
\end{tabular}
\label{Table2}
\caption{{\bf Double ionization  results  for Xe and He. }\hspace*{3.5cm}}
\end{table}
\end{center}
\vspace{0.cm}
\twocolumngrid
 
 \onecolumngrid

\begin{figure}
\centering
 \includegraphics[clip,height=1\textwidth]{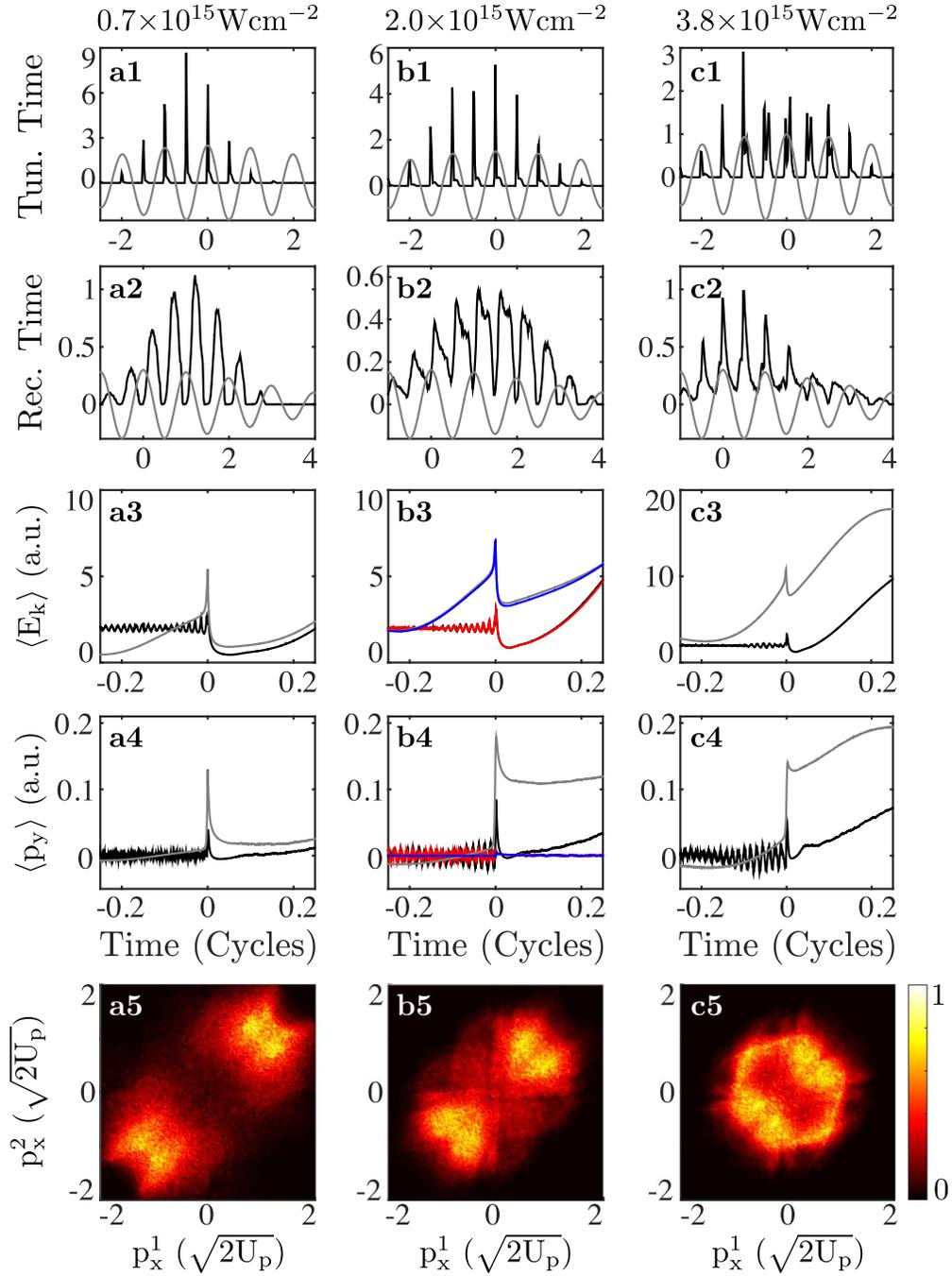}
 \caption{ {\bf Recollision underlying double ionization of He driven at 800 nm}. The intensities considered are 0.7$\times$10$^{15}$ Wcm$^{-2}$, denoted by (${\bf a}$), 2.0$\times$10$^{15}$ Wcm$^{-2}$, denoted by (${\bf b}$), and 3.8$\times$10$^{15}$ Wcm$^{-2}$, denoted by (${\bf c}$). The distribution of tunneling times (black line) is plotted in  (${\bf a1}$), (${\bf b1}$) and (${\bf c1}$).  The distribution of recollision  times (black line) is plotted in  (${\bf a2}$), (${\bf b2}$) and (${\bf c2}$). The electric field is denoted as a grey line in the plots of the tunneling and recollision times.  $\mathrm{\left<E_{k}^{1}\right>}$ (grey line) and  $\mathrm{\left<E_{k}^{2}\right>}$ (black line) are plotted as a function of time in (${\bf a3}$), (${\bf b3}$) and (${\bf c3}$) with time zero set equal to the recollision time of each double ionizing event. For 2.0$\times$10$^{15}$ Wcm$^{-2}$, $\mathrm{\left<E_{k}^{1}\right>}$ (blue line) and  $\mathrm{\left<E_{k}^{2}\right>}$ (red line) are also plotted in the absence of the magnetic field. $\mathrm{\left<p_{y}^{1}\right>}$ (grey line) and  $\mathrm{\left<p_{y}^{2}\right>}$ (black line) are plotted as a function of time in (${\bf a4}$), (${\bf b4}$) and (${\bf c4}$) with time zero set equal to the recollision time of each double ionizing event. For 2.0$\times$10$^{15}$ Wcm$^{-2}$, $\mathrm{\left<p_{y}^{1}\right>}$ (blue line) and  $\mathrm{\left<p_{2}^{2}\right>}$ (red line) are also plotted in the absence of the magnetic field. Correlated momenta along the direction of the electric field are plotted in (${\bf a5}$), (${\bf b5}$) and (${\bf c5}$).\hspace*{20cm}}
\label{figure4}
\end{figure}
\twocolumngrid

For single ionization  of He and Xe, it was shown that using the tunneling times as the starting point, the 3D-SMND with all Coulomb forces switched-off yields the correct order of magnitude for $\mathrm{\left <p_{y}^{1}\right>_{SI}}$. 
For double ionization of He and Xe, using the 3D-SMND model with all Coulomb forces switched-off  and with  initial conditions  taken to be the recollision times and velocities, $\mathrm{\left <p_{y}^{1}\right>_{DI}}$ and $\mathrm{\left <p_{y}
 ^{2}\right>_{DI}}$ are obtained and presented in Table 2. These values of  $\mathrm{\left <p_{y}^{1}\right>_{DI}}$ and $\mathrm{\left <p_{y}
 ^{2}\right>_{DI}}$ agree very well with the values obtained  using the 3D-SMND model  with all Coulomb forces accounted for, see Table 2. This agreement further supports that  recollision is the main factor determining $\mathrm{\left <p_{y}^{1}+p_{y}^{2} \right >_{DI}}$.

 Finally, in what follows, the  electron that contributes the most to the maximum value of $\mathrm{\left <p_{y}^1+p_{y}^{2} \right >_{DI}}$ is identified both for driven He and Xe. 
 $\mathrm{\left <p_{y}^{1}\right>_{DI}}$ and  $\mathrm{\left <p_{y}^{2}\right>_{DI}}$ are plotted as a function of time in \fig{figure4}, with  time zero set equal to the recollision time of each double ionizing trajectory.  It is shown in \fig{figure4}a4,b4,c4 that it is mainly $\mathrm{\left <p_{y}^{1}\right>_{DI}}$
   that changes  significantly at the recollision time. This change is more sharp for the smaller intensities  
  (\fig{figure4}a4,b4).  In \fig{figure4}b4, at intensity  2.0$\times$10$^{15}$ Wcm$^{-2}$, it is also  illustrated  that in the absence of the magnetic field both $\mathrm{\left <p_{y}^{1}\right>_{DI}}$ and  $\mathrm{\left <p_{y}^{2}\right>_{DI}}$ tend 
  to zero with time, as expected. In addition, 
  in \fig{figure3}c,d, for driven He and Xe, respectively, $\mathrm{\left <p_{y}^{1}\right>_{DI}}$ and  $\mathrm{\left <p_{y}^{2}\right>_{DI}}$ are  plotted as a function of intensity. It is seen that $\mathrm{\left <p_{y}^{1}\right>_{DI}}$  and $\mathrm{\left <p_{y}
  ^1+p_{y}^{2} \right >_{DI}}$ have maxima around the same intensities. At these intensities $\mathrm{\left <p_{y}^{1}\right>_{DI}}$ is 
  significantly larger than $\mathrm{\left <p_{y}^{2}\right>_{DI}}$. Thus,   $\mathrm{\left <p_{y}^{1}\right>_{DI}}$---the average momentum of the tunneling electron---is  the one affected the most by strong recollisions.

 \section{Discussion}
 
 It was shown that the average sum of the electron momenta along the propagation direction of the laser field has large values  at intensities where strong recollisions underlie double ionization. This is an unexpected result. For He driven by a near-infrared laser field and for Xe driven by a mid-infrared laser field, the intensities where  the average sum of the electron momenta along the propagation direction of the laser field is maximum are smaller than the intensities where magnetic field effects are predicted to be large. Thus, recollision probes magnetic field effects at smaller intensities than expected.  However, it can also be stated that a magnetic field probes strong recollisions through the measurement of the sum of the electron momenta along the propagation direction of the laser field. It is expected that the findings reported in this work will serve as a motivation for future studies. Such studies can identify, for instance, the effect the magnetic field has on  the different mechanisms of non-sequential double ionization, i.e. on direct and delayed double ionization and  the wavelengths where the magnetic field has the largest effect on recollisions.  

\section{Acknowledgments}A.E. is grateful for fruitful discussions with Paul Corkum. She  also acknowledges the EPSRC grant no. J0171831 and the use of the computational resources of Legion at UCL.

\section{Author contributions}
A.E. conceived the idea for the theoretical work, derived the theoretical framework and contributed many of the codes used for the computations. T. M. contributed some of the codes used for the computations, performed the computations and contributed to the analysis of the results.


\end{document}